\documentclass[conference]{IEEEtran}
\usepackage[T1]{fontenc} %
\usepackage[utf8]{inputenc} %
\usepackage[english]{babel} %
\usepackage[a4paper, total={184mm,239mm}]{geometry}
\usepackage{amsmath,amsthm,amsfonts}
\usepackage{colonequals} %
\usepackage[cmintegrals]{newtxmath} %
\usepackage{bm} %
\usepackage{comment} %
\usepackage{microtype}
\usepackage[
    backend=biber,
    style=ieee,
    sortlocale=en_US,
    sortcites=true,
    url=true,
    eprint=true,
    giveninits=false,
    dashed=false,
    minnames=1,
    maxnames=5
]{biblatex}
\AtEveryBibitem{%
    \clearname{editor}%
    \clearfield{series}%
    \clearfield{isbn}%
    \clearfield{issn}%
    \clearfield{day}%
    \clearfield{month}%
    \clearfield{place}%
    \clearlist{location}%
    \clearfield{doi}%
    \clearfield{pages}%
    \clearfield{volume}%
    \clearfield{number}%
    \ifentrytype{software}{%
    }{%
        \clearfield{url}%
        \clearfield{urlyear}%
    }%
}
\usepackage{csquotes} %
\usepackage[inline]{enumitem} %
\usepackage{tabularx} %
\usepackage{booktabs} %
\usepackage[hidelinks]{hyperref} %
\usepackage[capitalize,nameinlink]{cleveref}
\usepackage{xcolor}
\usepackage{graphicx}
\usepackage{braket}
\usepackage{stmaryrd}
\usepackage{mathtools}

\makeatletter
\long\def\@makecaption#1#2{%
    \ifx\@captype\@IEEEtablestring%
    \footnotesize\bgroup\par\centering\@IEEEtabletopskipstrut{\normalfont\footnotesize {#1.}\nobreakspace\scshape #2}\par\addvspace{0.5\baselineskip}\egroup%
    \@IEEEtablecaptionsepspace
    \else
        \@IEEEfigurecaptionsepspace
        \setbox\@tempboxa\hbox{\normalfont\footnotesize {#1.}\nobreakspace #2}%
        \ifdim \wd\@tempboxa >\hsize%
        \setbox\@tempboxa\hbox{\normalfont\footnotesize {#1.}\nobreakspace}%
        \parbox[t]{\hsize}{\normalfont\footnotesize \noindent\unhbox\@tempboxa#2}%
        \else%
            \hbox to\hsize{\normalfont\footnotesize\hfil\box\@tempboxa\hfil}%
        \fi\fi}
\makeatother
\makeatletter
\let\MYcaption\@makecaption
\makeatother
\usepackage{subcaption} %
\makeatletter
\let\@makecaption\MYcaption
\makeatother
\usepackage{wrapfig} %
\usepackage{float}  %
\usepackage{placeins} %
\usepackage{fontawesome5}
\faStyle{regular}
\usepackage{xspace}
\usepackage[ruled,vlined,linesnumbered]{algorithm2e}
\usepackage{fnpct} %
\usepackage{thmtools}
\declaretheoremstyle[
    spaceabove=-10pt,%
    spacebelow=-10pt,%
    headformat=\NAME\NUMBER \NOTE,%
    headpunct={.\vspace{-5pt}},%
    mdframed={},%
    bodyfont=\small%
]{constraintbox}
\declaretheorem[%
    name=C,%
    style=constraintbox,%
]{boxedconstraints}
\crefformat{boxedconstraints}{#2Box C#1#3}
\crefformat{boxedconstraints}{#2Box C#1#3}
\declaretheorem[%
    name=V,%
    style=constraintbox,%
]{boxedvariables}
\crefformat{boxedvariables}{#2Box V#1#3}
\crefformat{boxedvariables}{#2Box V#1#3}
\declaretheoremstyle[
    bodyfont=\itshape,%
    spaceabove=0pt,%
    spacebelow=0pt,%
]{example-style}
\declaretheorem[
    name=Example,%
    style=example-style,%
    numbered=unless unique,%
]{example}
\declaretheoremstyle[
    bodyfont=\itshape,%
    spaceabove=0pt,%
    spacebelow=0pt,%
]{problem-style}
\declaretheorem[
    name=Problem,%
    style=problem-style,%
    numbered=unless unique,%
]{problem}
\usepackage{siunitx}
\crefname{layout}{Layout}{Layouts}
\crefname{layout}{Layout}{Layouts}
\crefname{algocf}{Algorithm}{Algorithms}
\crefname{algocf}{Algorithm}{Algorithms}
\crefname{section}{Sec.}{Sec.}
\crefname{section}{Sec.}{Sec.}
\newcolumntype{R}{>{\raggedleft\arraybackslash}X}
\newcolumntype{C}{>{\centering\arraybackslash}X}

\definecolor{TUM_blue}{RGB}{0,101,189}
\colorlet{TUM_black}{black}
\colorlet{TUM_white}{white}
\definecolor{TUM_darkblue}{RGB}{0,82,147}
\colorlet{TUM_darkblue100}{TUM_darkblue}
\colorlet{TUM_darkblue80}{TUM_darkblue100!80}
\colorlet{TUM_darkblue50}{TUM_darkblue100!50}
\colorlet{TUM_darkblue20}{TUM_darkblue100!20}
\definecolor{TUM_verydarkblue}{RGB}{0,51,89}
\colorlet{TUM_verydarkblue100}{TUM_verydarkblue}
\colorlet{TUM_verydarkblue80}{TUM_verydarkblue100!80}
\colorlet{TUM_verydarkblue50}{TUM_verydarkblue100!50}
\colorlet{TUM_verydarkblue20}{TUM_verydarkblue100!20}
\colorlet{TUM_darkgrey}{TUM_black!80}
\colorlet{TUM_grey}{TUM_black!50}
\colorlet{TUM_lightgrey}{TUM_black!20}
\definecolor{TUM_beige}{RGB}{218,215,203}
\definecolor{TUM_orange}{RGB}{227,114,34}
\definecolor{TUM_green}{RGB}{162,173,0}
\definecolor{TUM_verylightblue}{RGB}{152,198,234}
\definecolor{TUM_lightblue}{RGB}{100,160,200}
\newcommand{\ie}{i.\,e.\nolinebreak\@\xspace}
\newcommand{\eg}{e.\,g.\nolinebreak\@\xspace}

\renewcommand{\phi}{\varphi}
\newcommand{\defeq}{\colonequals}
\newcommand{\abs}[1]{\left|#1\right|}

\newcommand{\ryd}[1]{e_{#1}}

\newcommand{\aod}[2]{a_{#1}^{\left(#2\right)}}
\newcommand{\xcoord}[2]{x_{#1}^{\left(#2\right)}}
\newcommand{\ycoord}[2]{y_{#1}^{\left(#2\right)}}
\newcommand{\row}[2]{r_{#1}^{\left(#2\right)}}
\newcommand{\column}[2]{c_{#1}^{\left(#2\right)}}
\newcommand{\horizontalOffset}[2]{h_{#1}^{\left(#2\right)}}
\newcommand{\verticalOffset}[2]{v_{#1}^{\left(#2\right)}}
\newcommand{\loadColumn}[2]{{}^cl_{#1}^{\left(#2\right)}}
\newcommand{\loadRow}[2]{{}^rl_{#1}^{\left(#2\right)}}
\newcommand{\storeColumn}[2]{{}^cs_{#1}^{\left(#2\right)}}
\newcommand{\storeRow}[2]{{}^rs_{#1}^{\left(#2\right)}}
\newcommand{\maxX}{\mathsf{X_{max}}}
\newcommand{\maxY}{\mathsf{Y_{max}}}
\newcommand{\maxC}{\mathsf{C_{max}}}
\newcommand{\maxR}{\mathsf{R_{max}}}
\newcommand{\maxH}{\mathsf{H_{max}}}
\newcommand{\maxV}{\mathsf{V_{max}}}
\newcommand{\maxG}{\mathsf{G}}
\newcommand{\minEntanglingZone}{\mathsf{E_{min}}}
\newcommand{\maxEntanglingZone}{\mathsf{E_{max}}}
\newcommand{\gateIndices}{\left[\maxG\right]}
\newcommand{\maxInteractionRadius}{\mathsf{r}}
\newcommand{\maxStages}{\mathsf{S}}
\newcommand{\image}[1]{\mathfrak{im}\left(#1\right)}
\renewcommand{\implies}{\rightarrow}
\renewcommand{\equiv}{\leftrightarrow}

\title{Optimal State Preparation for Logical Arrays \\on Zoned Neutral Atom Quantum Computers\vspace{-8pt}}

\author{
    \IEEEauthorblockN{
        Yannick Stade\IEEEauthorrefmark{1},
        Ludwig Schmid\IEEEauthorrefmark{1},
        Lukas Burgholzer\IEEEauthorrefmark{1},
        and Robert Wille\IEEEauthorrefmark{1}\IEEEauthorrefmark{2}
    }
    \IEEEauthorblockA{\IEEEauthorrefmark{1}%
    Chair for Design Automation,
        Technical University of Munich,
        Munich, Germany
    }
    \IEEEauthorblockA{\IEEEauthorrefmark{2}%
    Software Competence Center Hagenberg GmbH,
        Hagenberg, Austria
    }
    \{%
    yannick.stade,%
    ludwig.s.schmid,%
    lukas.burgholzer,%
    robert.wille%
    \}@tum.de\\
    \href{https://www.cda.cit.tum.de/research/quantum}{www.cda.cit.tum.de/research/quantum}
    \vspace{-10pt}
}
\hypersetup{ %
    pdftitle={Optimal State Preparation for Logical Arrays on Zoned Neutral Atom Quantum Computers},
    pdfsubject={Design, Automation and Test in Europe Conference 2025},
    pdfauthor={
        Yannick Stade,
        Ludwig Schmid,
        Lukas Burgholzer,
        Robert Wille
    }
}

\begin{document}
\maketitle
\begin{abstract}
Quantum computing promises to solve problems previously deemed infeasible.
However, high error rates necessitate quantum error correction for practical applications.
Seminal experiments with zoned neutral atom architectures have shown remarkable potential for fault-tolerant quantum computing.
To fully harness their potential, efficient software solutions are vital.
A key aspect of quantum error correction is the initialization of physical qubits representing a logical qubit in a highly entangled state.
This process, known as state preparation, is the foundation of most quantum error correction codes and, hence, a crucial step towards \mbox{fault-tolerant} quantum computing.
Generating a schedule of target-specific instructions to perform the state preparation is highly complex.
First software tools exist but are not suitable for the zoned neutral atom architectures.
This work addresses this gap by leveraging the computational power of SMT solvers and generating minimal schedules for the state preparation of logical arrays.
Experimental evaluations demonstrate that actively utilizing zones to shield idling qubits consistently results in higher fidelities than solutions disregarding these zones.
The complete code is publicly available in open-source as part of the \emph{Munich Quantum Toolkit}~(MQT) at \url{https://github.com/cda-tum/mqt-qmap/tree/main/src/na/nasp}.
\end{abstract}

\section{Introduction}\label{sec:introduction}

Quantum computing has emerged as a promising technology for solving computationally intensive tasks~\cite{preskillQuantumComputingNISQ2018}.
However, one major challenge is the high error rates in quantum devices.
Although further advancements in physical implementations are expected to reduce these error rates, it is widely understood that practical quantum computations must always be \mbox{fault-tolerant}~\cite{campbellRoadsFaulttolerantUniversal2017}.
To this end, we can leverage the well-developed theory of \emph{Quantum Error Correction}~(QEC,~\cite{shorFaulttolerantQuantumComputation1996,gottesmanStabilizerCodesQuantum1997,bombinGaugeColorCodes2015,steaneFastFaulttolerantFiltering2004}).

In QEC, a \emph{logical qubit}, \ie, a qubit on the algorithmic level, is redundantly encoded into an array of \emph{physical qubits}, a so-called \emph{logical array}.
One crucial part of QEC is preparing the physical qubits in a shared state representing the initial state of the encoded logical qubit before the actual algorithm can be executed.
This process is known as \emph{state preparation}.
For many QEC codes, a circuit to perform the state preparation can be generated~\cite{zenQuantumCircuitDiscovery2024} and has then to be translated into a schedule of target-specific instructions.

By design, implementing QEC requires significantly more physical qubits than qubits at the logical level.
However, as quantum devices continue to scale up in their qubit count~\cite{manetschTweezerArray61002024}, the application of QEC is finally becoming increasingly feasible.
In experiments, researchers have already demonstrated the preparation of 40 logical qubits in the \(|0\rangle_L\)–state comprising 280 physical qubits using a \emph{zoned neutral atom architecture}, where different operations are performed in designated spatially separated zones~\cite{bluvsteinLogicalQuantumProcessor2023}.
Consequently, we urgently need appropriate software tools that automate the task of translating a state preparation circuit into a schedule of instructions specific to this quantum device.

For this highly complex task, we propose to use \emph{Satisfiability Modulo Theories}~(SMT) solvers~\cite{demouraZ3EfficientSMT2008}---a powerful and established method in traditional design automation---and apply them to the problem of compiling for zoned neutral atom architectures.
The resulting approach allows the generation of \emph{minimal} schedules in terms of the number of error-prone and time-intensive operations---crucial to achieving high fidelity.

Experimental evaluations show that the proposed approach consistently achieves higher fidelities than solutions that do not exploit the benefits of the zoned architecture.
We extend those results by exploring different variations of the existing zoned neutral atom architecture, which provides valuable insights for the design of future quantum devices.
The full code of the proposed approach is publicly available as part of the \emph{Munich Quantum Toolkit}~(MQT,~\cite{mqt}) at \url{https://github.com/cda-tum/mqt-qmap/tree/main/src/na/nasp}.

\section{Background}\label{sec:preliminaries}

In order to keep the paper self-contained, this section provides
\begin{enumerate*}[label=(\arabic*)]
\item a brief review of the basics of QEC with a focus on state preparation as well as
\item an overview of neutral atom architectures.
\end{enumerate*}

\begin{figure*}[tb]
\centering
\includegraphics[width=\linewidth]{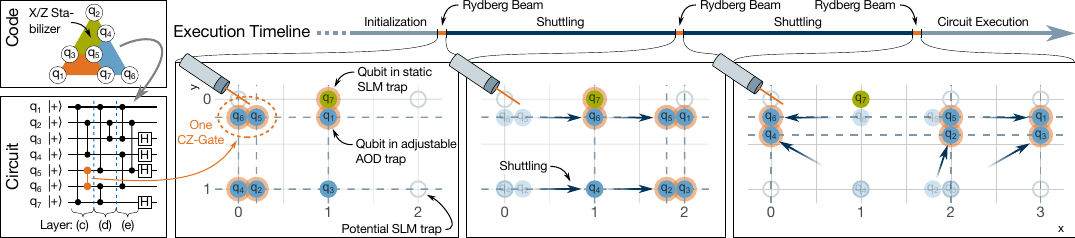}\\
\raggedright\vspace{-89pt}\hspace{3pt}%
\begin{subfigure}{78pt}
\caption{\raggedright}
\label{subfig:motivation code}
\end{subfigure}\\
\vspace{58pt}\hspace{3pt}%
\begin{subfigure}{78pt}
\caption{\raggedright}
\label{subfig:motivation circuit}
\end{subfigure}%
\hspace{7pt}%
\begin{subfigure}{135pt}
\caption{\raggedright}
\label{subfig:motivation 1}
\end{subfigure}%
\hspace{7pt}%
\begin{subfigure}{123pt}
\caption{\raggedright}
\label{subfig:motivation 2}
\end{subfigure}%
\hspace{7pt}%
\begin{subfigure}{156pt}
\caption{\raggedright}
\label{subfig:motivation 3}
\end{subfigure}%
\vspace{-4pt}
\caption{
The visualization of the stabilizers of the Steane code (\cref{subfig:motivation code}) and a quantum circuit preparing the logical \(|0\rangle_\mathrm{L}\)-state for this code (\cref{subfig:motivation circuit}).
In the remaining part, a sequence of three Rydberg beams with intermediate shuttling indicated by the dark blue arrows realizing the circuit on the left.
Each frame (\cref{subfig:motivation 1,subfig:motivation 2,subfig:motivation 3}) corresponds to one layer of CZ-gates (separated by the barriers) in the circuit on the left.
}
\label{fig:motivation}
\vspace{-8pt}
\end{figure*}

\subsection{Quantum Error Correction and State Preparation}\label{subsec:qec}
Qubits and quantum operations are inherently noisy due to their susceptibility to environmental influences and control errors.
In order to prevent error accumulation and to enable large-scale quantum computations, several QEC protocols have been developed over the past decades~\cite{shorFaulttolerantQuantumComputation1996,gottesmanStabilizerCodesQuantum1997,bombinGaugeColorCodes2015,steaneFastFaulttolerantFiltering2004}.
Similar to classical error correction, the fundamental principle of QEC is to redundantly encode the information of one or more \emph{logical qubits} into an array of \emph{physical qubits}, a so-called \emph{logical array}.
A code that requires \(n\) physical qubits to encode \(k\) logical qubits with \emph{code distance} \(d\)---the minimum number of physical errors that can lead to a logical error---is denoted as an \mbox{$\llbracket n, k, d \rrbracket$ code}.

The most common QEC codes are \emph{Stabilizer Codes}~\cite{gottesmanStabilizerCodesQuantum1997}.
Here, the logical qubit is encoded in the \mbox{$+1$-eigenspace} of a set of $n-k$ linear independent Pauli operators (\emph{stabilizers}) of the form $P^{\otimes n}$, where each entry $P \in \set{I,X,Y,Z}$ represents a Pauli operator acting on a single qubit.
This constraint reduces the $2^n$ dimensional Hilbert space to the $2^k$ dimensional \emph{codespace} for the logical qubits.

\begin{example}
\cref{subfig:motivation code} illustrates the Steane code, a $\llbracket 7, 1, 3 \rrbracket$ code, which is the smallest instance of a 2D color code~\cite{bombinGaugeColorCodes2015}.
Circles represent physical qubits, and each of the three facets (green, blue, orange) defines two stabilizers: One with \mbox{Z-operators} and one with \mbox{X-operators}, acting on all qubits at the vertices of the facet, while identity operators act on the rest.
This results, \eg, in the stabilizers \(IZZZZII\) and \(IXXXXII\) for the green facet.
\end{example}

Using QEC, the state $\ket{\psi}_L$ of a logical qubit can be repeatedly checked for errors and corrected involving \emph{syndrome extraction}, \emph{decoding}, and \emph{Pauli frame tracking}~\mbox{\cite{gottesmanStabilizerCodesQuantum1997,roffeQuantumErrorCorrection2019}}.
However, before correction or computation, the state must be prepared as a valid logical state within the codespace.
This process, known as \emph{state preparation} (or \emph{logical encoding}), involves a \emph{state preparation circuit} that operates on the physical qubits.

Various methods have been proposed to design efficient state preparation circuits for different QEC codes.
These include optimal schemes%
~\cite{pehamAutomatedSynthesisFault2024}, heuristic approaches such as the "Latin rectangle" method~\cite{steaneFastFaulttolerantFiltering2004}, and machine learning techniques based on reinforcement learning~\cite{zenQuantumCircuitDiscovery2024}.

\begin{example}
\cref{subfig:motivation circuit} shows a state preparation circuit for the Steane code.
The physical qubits are initialized in the $\ket{+}$ state followed by a series of CZ-gates and Hadamard gates at the end on selected qubits.
\end{example}

\subsection{Neutral Atom Architecture}\label{subsec:na}
Over time \emph{neutral atoms}~\mbox{\cite{saffmanQuantumComputingNeutral2019,henrietQuantumComputingNeutral2020,winterspergerNeutralAtomQuantum2023,schmidComputationalCapabilitiesCompiler2023,bluvsteinQuantumProcessorBased2022}} have emerged as a promising platform for large-scale, fault-tolerant quantum computations~\cite{bluvsteinLogicalQuantumProcessor2023}.
Here, qubits are encoded in the electronic states of individual neutral atoms such as Rb, Sr, or Yb, which are trapped in optical tweezers and laser-cooled to their motional ground state~\cite{barredoAtombyatomAssemblerDefectfree2016}.

Single-qubit gates are realized as state transitions driven by global and local lasers~\cite{saffmanQuantumComputingNeutral2019,bluvsteinQuantumProcessorBased2022}.
\mbox{Multi-qubit} gates are implemented using the Rydberg blockade%
~\mbox{\cite{everedHighfidelityParallelEntangling2023,bluvsteinQuantumProcessorBased2022}}.
Global \emph{Rydberg beams} illuminate the entire trap area, enabling multiple parallel CZ-gates between all adjacent qubits.
Isolated atoms, however, undergo an imperfect identity operation as they are still excited to the Rydberg state, suffering from Rydberg decay, which is a major error source for CZ-gates~\cite{everedHighfidelityParallelEntangling2023}.

\begin{example}
\cref{subfig:motivation 1} illustrates the application of a global Rydberg beam, performing the first three parallel CZ-gates of the circuit in \cref{subfig:motivation circuit}, \ie, between the pairs of qubits $(q_1,q_7), (q_6,q_5),$ and $(q_4,q_2)$.
Qubit $q_3$ is isolated and does not participate in any \mbox{CZ-gate}. %
\end{example}

\begin{figure*}[!tp]
\centering
\includegraphics[width=\linewidth]{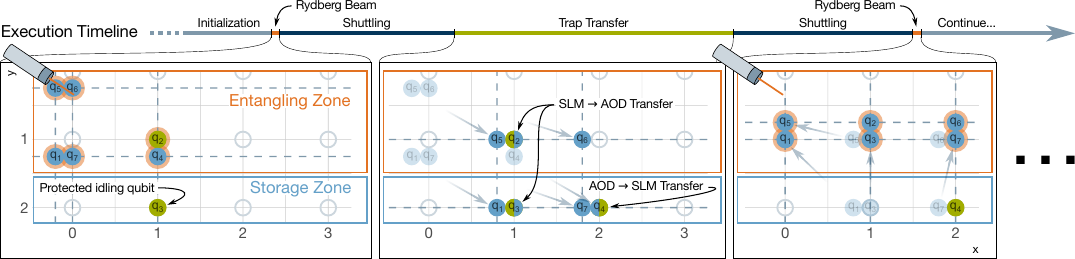}\\
\raggedright\vspace{-16pt}\hspace{3pt}%
\begin{subfigure}{177.5pt}
\caption{\raggedright}
\label{subfig:zone 1}
\end{subfigure}%
\hspace{7pt}%
\begin{subfigure}{165.5pt}
\caption{\raggedright}
\label{subfig:zone 2}
\end{subfigure}%
\hspace{7pt}%
\begin{subfigure}{120pt}
\caption{\raggedright}
\label{subfig:zone 3}
\end{subfigure}%
\vspace{-4pt}%
\caption{
The first three steps of a schedule executing the same circuit as in \cref{subfig:motivation circuit} on a zoned neutral atom architecture.
Throughout the two Rydberg beams shown in \cref{subfig:zone 1,subfig:zone 3}, qubits \(q_3\) and \(q_4\), respectively, are protected in the storage zone.
\cref{subfig:zone 2} shows the necessary trap transfer.
}
\label{fig:zone}
\vspace{-8pt}
\end{figure*}

For the application of \mbox{multi-qubit} gates, their operands must be brought in proximity to each other.
This is achieved by rearranging the atoms during computation---allowing for arbitrary qubit connectivity~\cite{bluvsteinQuantumProcessorBased2022}.
Experimentally, this is realized using two types of optical traps:
\begin{enumerate*}[label=(\arabic*)]
\item \emph{Spatial Light Modulator}~(SLM) traps, which define single static trap sites and
\item \emph{Acousto-Optic Deflector}~(AOD) traps, which create adjustable trap sites at crossings of perpendicularly intersecting AOD lines~\mbox{\cite{beugnonTwodimensionalTransportTransfer2007,bluvsteinQuantumProcessorBased2022}}.
\end{enumerate*}
An atom can be transferred from SLM to AOD traps and vice versa---called \emph{loading} and \emph{storing}, respectively---by activating or deactivating one AOD line crossing the atom~\cite{beugnonTwodimensionalTransportTransfer2007,bluvsteinQuantumProcessorBased2022}.
All atoms on the (de-)activated AOD line perform the same operation simultaneously, and unintended operations on other qubits must be carefully avoided~\cite{bluvsteinQuantumProcessorBased2022,stadeAbstractModelEfficient2024}.
Loaded atoms are shuttled by moving AOD rows and columns whose order must be preserved~\cite{schmidComputationalCapabilitiesCompiler2023}.

\begin{example}
In \cref{subfig:motivation 1}, qubit \(q_7\) is placed in an SLM trap (green), while the other qubits $q_1, \dots, q_6$ are loaded in AOD traps (blue) in a 2D grid formed by two rows and three columns (dashed lines).
Here, qubits \(q_4, q_2\), and \(q_3\) share the same AOD row.
Turning off this row would store all three qubits back into SLM traps.
In \cref{subfig:motivation 2}, the AOD columns are moved to the right, shuttling the AOD qubits (blue) indicated by the dark blue arrows relative to the static SLM qubit \(q_7\) (green).
In total, the steps in \cref{subfig:motivation 1,subfig:motivation 2,subfig:motivation 3} execute the state preparation circuit for the Steane code, depicted in \cref{subfig:motivation circuit}. %
\end{example}

In order to execute entire quantum circuits, multiple steps of Rydberg beams and shuttling---potentially involving trap transfers---must take turns.
Finding minimal schedules is crucial to compiling for neutral atom architectures~\cite{schmidComputationalCapabilitiesCompiler2023}.
Several tools have been proposed, including optimal solutions~\cite{brandhoferOptimalMappingNearTerm2021,tanCompilingQuantumCircuits2024}, scalable heuristics~\cite{tanCompilationDynamicallyFieldProgrammable2024,bakerExploitingLongDistanceInteractions2021,wangFPQACCompilationFramework2023}, and combinations of shuttling with alternative routing methods~\cite{schmidHybridCircuitMapping2023,nottinghamDecomposingRoutingQuantum2023}.
However, they all lack support for the novel \emph{zoned} neutral atom architectures. %

\section{Considered Problem}\label{sec:motivation}
A huge problem with \mbox{state-of-the-art} \mbox{state-preparation} for neutral atoms is that the Rydberg beams applied to execute \mbox{CZ-gates} affect not only the desired qubits but also the idling qubits.
For example, during the step, illustrated in \cref{subfig:motivation 1}, qubit~\(q_3\)
is illuminated by the Rydberg beam even though it is not involved in any CZ-gate.
In order to avoid this unnecessary error source by the faulty identity operation, the Rydberg beam should only illuminate qubits that undergo a \mbox{CZ-gate}.
This is where \emph{zoned} neutral atom architectures~\cite{bluvsteinLogicalQuantumProcessor2023} come into play.

Here, the architecture is divided into spatially separated zones of three kinds:
\begin{enumerate*}[label=(\arabic*)]
\item \emph{entangling zones}, where a global Rydberg beam can perform (entangling) CZ-gates between pairs of adjacent qubits,
\item \emph{storage zones}, in which qubits are shielded from interferences with the Rydberg beam and
where the qubits can be kept for a long time with very high coherence (single-qubit gates can be performed here as well), and
\item \emph{readout zones}, in which selected qubits can be measured without collapsing the state of qubits in other zones.\footnote{Throughout the rest of this work, we will only consider storage and entangling zones. The treatment of readout zones can be handled analogously.}
\end{enumerate*}

\begin{example}
\cref{subfig:zone 1} shows a schematic drawing of a zoned neutral atom architecture.
While applying the Rydberg beam, qubit \(q_3\) is shielded in the storage zone.
For the next Rydberg beam as in \cref{subfig:zone 3}, the qubits \(q_3\) and \(q_4\) have to be swapped.
This swapping is no longer possible with shuttling alone and requires trap transfers as illustrated in \cref{subfig:zone 2}, where qubit \(q_4\) is stored in the storage zone and qubits \(q_3\) and \(q_2\) are loaded.
\end{example}
The required trap transfers take about 100 times longer than other operations on this %
architecture~\cite{bluvsteinLogicalQuantumProcessor2023,everedHighfidelityParallelEntangling2023}.
Hence, their use should be minimized to the absolute minimum.

Both aspects–––the consideration of different zones and the demand-driven use of transfers–––are challenges not supported by existing tools reviewed in \cref{sec:preliminaries}.
Those tools only assume a single zone (that corresponds to the entangling zone), including all unavoidable error sources.
An exception are the compilers proposed in~\cite{deckerArcticFieldProgrammable2024,stadeAbstractModelEfficient2024}, which can handle zoned architectures.
However, the former cannot lead to minimal schedules as qubits are moved outside the entangling zone after every operation.
The latter is designed for routing entire logical arrays and assumes that these arrays (representing logical qubits) have already been prepared.
As such, the crucial state preparation step is missing.

In this work, we generate minimal schedules of target-specific instructions to perform the state preparation of logical arrays indispensable for most QEC codes.
The circuits for the state preparation of arbitrary stabilizer codes can be expressed following the same structure as the circuit in \cref{subfig:motivation circuit}, \ie, first, the physical qubits are prepared in the \(|+\rangle\)-state, then a set of \mbox{CZ-gates} create a so-called \emph{graph state}, and finally, the logical qubit is prepared in the \mbox{\(|0\rangle_L\)-state} by applying Hadamard gates on selected qubits~\cite{amaroScalableCharacterizationLocalizable2020}.
Since the preparation of the physical qubits in the \(|+\rangle\)-state, as well as the Hadamard gates at the end, can be realized everywhere on the architecture by rotational gates and does not require shuttling, this gives rise to the following problem statement:

\begin{problem}
Given a set of \mbox{CZ-gates}, find a schedule of Rydberg beams, trap transfers, and shuttling operations that realizes the given \mbox{CZ-gates} on the zoned neutral atom architecture with high fidelity.
\end{problem}

\section{Proposed Solution}\label{sec:solution}
In this work, we tackle the complexity of the problem reviewed above by utilizing the computational power of SMT solvers~\cite{demouraZ3EfficientSMT2008}.
They are a powerful and established method from the conventional design automation realm and, as we show in the following, are well-suited for tackling problems arising in the compilation of quantum computing. %
The main idea is as follows: First, all possible schedules are represented symbolically, \ie, through a set of variables that represents, \eg, qubit positions, shuttling operations, and gate \mbox{execution---leading} to a \emph{symbolic formulation} representing the entire search space.
Then, constraints and an objective function are employed to ensure
\begin{enumerate*}[label=(\arabic*)]
\item those variables can only assume values corresponding to valid solutions and\label{itm:valid}
\item that the solution is optimal\label{itm:optimal}
\end{enumerate*}.
Finally, the resulting formulation is passed to an SMT solver trying to determine such an assignment.
Naturally, the complexity of the problem remains, but as shown later in \cref{sec:evaluation}, the proposed solution still facilitates the creation of schedules for practical instances.
Even if the solver fails due to the complexity, it may still produce a valid assignment (cf.~\ref{itm:valid}), albeit not an optimal one (cf.~\ref{itm:optimal}).
From this assignment, the desired schedule can eventually be derived.
The following sections describe the construction of the symbolic formulation, the constraints, and the objective function that encode the problem under consideration.

\subsection{Symbolic Formulation}\label{subsec:variables}
To ease the proposed formalization, we divide the time into discrete stages, and the space into finitely many trap sites, inspired by the model proposed in~\cite{tanCompilingQuantumCircuits2024}.
Those trap sites are grouped hierarchically into \emph{interaction sites}.
Each interaction site comprises one SLM trap in its center and potential AOD traps around it.
At the beginning of a stage, each qubit is in one of these traps.
Hence, its position is defined by the x- and y-coordinates of its interaction site together with the horizontal and vertical offset from the center within the interaction site.
Additionally, to allow only as many AOD columns and rows as the architecture can provide (and to facilitate the verification of AOD constraints~\cite{stadeAbstractModelEfficient2024}), qubits in AOD traps are assigned to their respective AOD column and row.
For the rest of the paper, the letter \(q\) refers to a qubit, and the letter \(t\) denotes a stage.
In the formulas, \(\maxX\), \(\maxY\) denote the max. x- and y-coordinates, \(\maxH\), \(\maxV\) the max. horizontal and vertical offsets, and \(\maxC\), \(\maxR\) the number of AOD columns and rows, respectively.

\begin{boxedvariables}[Positioning Qubits]
\label{con:qubit variables}
\begin{align}
& \forall_{q,t}\: 0 \le \xcoord{t}{q} \le \maxX, \quad 0 \le \ycoord{t}{q} \le \maxY \label{eq:position vars}                 \\
& \forall_{q,t}\: \abs{\horizontalOffset{t}{q}} \le \maxH, \quad \abs{\verticalOffset{t}{q}} \le \maxV \label{eq:offset vars} \\
& \forall_{q,t}\: \aod{t}{q} \in \set{\mathsf{true}, \mathsf{false}} \label{eq:aod var}                                       \\
& \forall_{q,t}\: 0 \le \column{t}{q} \le \maxC, \quad 0 \le \row{t}{q} \le \maxR \label{eq:aod columns/rows}
\end{align}
\end{boxedvariables}

\cref{eq:position vars,eq:offset vars} ensure that the x- and y-coordinates and the offset for each qubit are within the bounds of the architecture.
The variable in \cref{eq:aod var} determines whether a qubit is loaded in an AOD trap; \cref{eq:aod columns/rows} limits the number of available AOD columns and rows.

\begin{figure}[!tp]
\centering
\includegraphics[width=\linewidth]{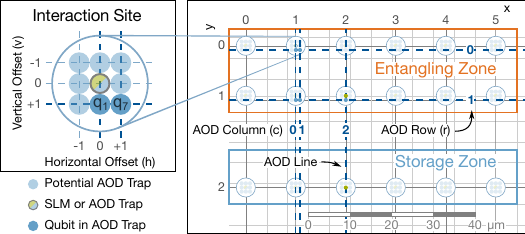}\\
\raggedright\vspace{-47pt}\hspace{3pt}%
\begin{subfigure}{84pt}
\caption{\raggedright}
\label{subfig:interaction site}
\end{subfigure}\\
\vspace{15.5pt}
\hspace{94pt}%
\begin{subfigure}{158pt}
\caption{\raggedright}
\label{subfig:architecture}
\end{subfigure}%
\vspace{-4pt}%
\caption{
Schematic zoned neutral atom architecture with one storage zone corresponding to \(\maxX = 5, \maxY = 2, \maxH = \maxV = 1, \maxC = 2, \maxR = 1\) and \(\minEntanglingZone = 0, \maxEntanglingZone = 1\).
In the center of each interaction site is an SLM trap surrounded by potential AOD traps.
}
\label{fig:interaction site}
\vspace{-8pt}
\end{figure}

\begin{example}
\cref{subfig:architecture} illustrates a zoned architecture with an entangling zone and storage zone.
A coordinate system spans the plane, identifying each interaction site with its x- and y-coordinates.
\cref{subfig:interaction site} details the interaction site at \(x=1\) and \(y=0\), showing the trap sites' possible horizontal and vertical offsets.
Hence, for the qubit \(q_7\) this results, \eg, in the following variable assignment at stage \(t=1\): \(\ycoord{1}{q_7}\!\! = 0, \xcoord{1}{q_7}\!\! = \horizontalOffset{1}{q_7}\!\! = \verticalOffset{1}{q_7}\!\! = 1\).
In \cref{subfig:architecture}, AOD lines (dashed blue lines) carry their index as a label, resulting in \(\column{1}{q_7} = 1, \row{1}{q_7} = 0\), and \(\aod{1}{q_7}=\mathtt{true}\).%
\end{example}

There are two different kinds of stages: \emph{Execution stages} and \emph{transfer stages}.
An execution stage starts with a Rydberg beam that causes CZ-gates between adjacent qubits, followed by shuttling the qubits to their position in the next stage.
At the beginning of a transfer stage, qubits can transfer from static SLM to adjustable AOD traps and vice versa, followed by the shuttling of qubits to their subsequent position.

We assume that the CZ-gates constituting the state preparation circuit are given as a list \(\langle g_1, \dots, g_\maxG \rangle\) of (unordered) pairs of qubits.
Later, we need to refer to the (unordered) pair of operands itself, which we denote as \(\image{g_i}\) for some \(i\in\gateIndices\), where \(\gateIndices\defeq\set{1,\dots,\maxG}\).

\begin{boxedvariables}[Executing Gates]\hfil\\
\label{con:gate vars}
\begin{minipage}{.45\linewidth}
\begin{align}
& \forall_{i\in\gateIndices}\: 0 \le g_i < \maxStages
\label{eq:gate vars}
\end{align}
\end{minipage}
\hfill
\begin{minipage}{.45\linewidth}
\begin{align}
& \forall_{t}\: \ryd{t} \in \set{\mathsf{true}, \mathsf{false}} \label{eq:stage vars}
\end{align}
\end{minipage}
\end{boxedvariables}

The constant \(\maxStages\) is an input parameter and denotes the total number of stages, and hence \cref{eq:gate vars} ensures that all gates are executed within the given number of stages.
The variable in \cref{eq:stage vars} determines whether a stage is an execution stage or not.

\begin{example}
Note that \cref{fig:interaction site} extends the step depicted in \cref{subfig:motivation 1}.
The corresponding stage is an execution stage where the first three CZ-gates of the circuit in~\cref{subfig:motivation circuit} are executed, \ie, \(g_i = 1\) if \(i=1,2,3\) else \(g_i\neq1\) and \(\ryd{1} = \mathtt{true}\).
\end{example}

Qubits can be transferred from static SLM to adjustable AOD traps and vice versa, which is called \emph{loading} and \emph{storing}, respectively.
To avoid unintended load and store operations, we label every AOD column and row that is responsible for a load or store operation accordingly. %

\begin{boxedvariables}[Loading/Storing Qubits]
\label{con:transfer vars}
\begin{align}
& \forall_{k\in\set{0,\dots,\maxC}, t}\: \loadColumn{t}{k}, \storeColumn{t}{k} \in \set{\mathsf{true}, \mathsf{false}} \label{eq:transfer column var} \\
& \forall_{k\in\set{0,\dots,\maxR}, t}\: \loadRow{t}{k}, \storeRow{t}{k} \in \set{\mathsf{true}, \mathsf{false}} \label{eq:transfer row var}
\end{align}
\end{boxedvariables}

In \cref{eq:transfer column var}, when a variable has the value \texttt{true} for a column, then all qubits in this column must be loaded or stored (in that stage), respectively; the same holds for rows in \cref{eq:transfer row var}, analogously.

\begin{example}
For stage \(t=2\), depicted in \cref{subfig:motivation 2}, \(\loadColumn{2}{1}\) and \(\storeColumn{2}{2}\) are \texttt{true} whereas all other variables from \cref{con:transfer vars} are \texttt{false}.
This reflects that all qubits previously on column 1 are stored, and qubits afterward on column~2 are loaded.
\end{example}

\subsection{Constraints}\label{subsec:constraints}
Most of the possible assignments are invalid because they describe physically infeasible schedules.
To limit the search space to assignments that describe valid solutions, we introduce constraints on the variables to reflect the restrictions imposed by the neutral atom architecture.
Most obviously, a trap can hold at most one qubit at a time.

\begin{boxedconstraints}[Positioning Qubits]
\label{con:trap constraints}
\begin{align}
& \begin{aligned}
& \forall_{\substack{t,q,q'                                                                                              \\q\neq q}}\: \left( \horizontalOffset{t}{q} = \horizontalOffset{t}{q'} \right) \land \left( \verticalOffset{t}{q} = \verticalOffset{t}{q'} \right) \\[-1.5ex]
& \qquad \implies \left( \xcoord{t}{q} \neq \xcoord{t}{q'} \right) \lor \left( \ycoord{t}{q} \neq \ycoord{t}{q'} \right)
\end{aligned}\label{eq:trap occupancy} \\
& \forall_{q, t}\: \lnot\aod{t}{q} \implies \left( \horizontalOffset{t}{q} = 0 = \verticalOffset{t}{q} \right) \label{eq:slm center}
\end{align}
\end{boxedconstraints}

\cref{eq:trap occupancy} requires qubits with identical horizontal and vertical offsets to be located at different interaction sites, implying that two qubits can never be at the same position.
As only the center qubit in an interaction site can be confined in an SLM trap, such a qubit must have zero offsets, cf.~\cref{eq:slm center}.

AOD traps are created at crossings of AOD columns with rows.
For a consistent model, we arrange the columns from left to right in increasing order.
Analogously, there is a constraint for the rows, which is omitted for brevity.

\begin{boxedconstraints}[Ordering AOD lines]
\label{con:aod constraints}
\begin{align}
& \begin{aligned}
& \forall_{q,q',t}\: \left( \aod{t}{q} \land \aod{t}{q'} \right) \implies \Big(\column{t}{q} < \column{t}{q'} \equiv                                          \\
& \qquad\quad \xcoord{t}{q} < \xcoord{t}{q'} \lor \left(\xcoord{t}{q} = \xcoord{t}{q'} \land \horizontalOffset{t}{q} < \horizontalOffset{t}{q'}\right)\!\Big)
\end{aligned} \label{eq:AOD column order} %
\end{align}
\end{boxedconstraints}

At the beginning of an execution stage, all qubits in the entangling zone in proximity to each other undergo a CZ-gate.
Central to our work is the utilization of the different zones provided by the architecture.
Idling qubits are moved outside the entangling zone to be shielded from the Rydberg beam.

\begin{boxedconstraints}[Executing Gates]
\label{con:gate execution constraints}
\begin{align}
& \begin{aligned}
& \forall\!\!\!\!{}_{\substack{i\in\gateIndices                                                                                                                                                                        \\g_i=\left(q,q'\right)}}\!\! \left( g_i = t \right) \implies \Big( \ryd{t} \land \left( \xcoord{t}{q}\! = \xcoord{t}{q'} \right) \land \left( \ycoord{t}{q}\! = \ycoord{t}{q'} \right) \\[-1.5ex]
& \qquad\qquad \land \left( \abs{\horizontalOffset{t}{q'} - \horizontalOffset{t}{q}} < \maxInteractionRadius \right) \land \left( \abs{\verticalOffset{t}{q'} - \verticalOffset{t}{q}} < \maxInteractionRadius \right) \\
& \quad \land \left( \minEntanglingZone \le \ycoord{t}{q}\! \le \maxEntanglingZone \right) \land \left( \minEntanglingZone \le \ycoord{t}{q'}\! \le \maxEntanglingZone \right)\!\Big)
\end{aligned} \label{eq:gate execution} \\
& \forall_{\substack{i,j\in\gateIndices                                                                                                                                                                                                                                                                                                                                                     \\i\neq j}}\: \left(g_i \neq g_j\right) \qquad\text{if}\quad \image{g_i} \cap \image{g_j} \neq \emptyset \label{eq:gates on same qubit} \\[-1ex]
& \forall_{q}\: \ryd{t} \implies \lnot\Big(\Big(\forall\!\!{}_{\substack{i\in\gateIndices                                                                                                                                                                                                                                                                                                   \\q\in\image{g_i}}} g_i \neq t\Big) \land \Big(\minEntanglingZone \le \ycoord{t}{q} \le \maxEntanglingZone \Big)\Big) \label{eq:shield idle qubits}
\end{align}
\end{boxedconstraints}

For a gate \(g_i\) to be executed in stage \(t\), \cref{eq:gate execution} ensures that all prerequisites are fulfilled such that the gate is physically executed.
Gates involving the same qubits cannot be executed simultaneously, cf.~\cref{eq:gates on same qubit}.
The constants \(\minEntanglingZone\) and \(\maxEntanglingZone\) are the bounds of the entangling zone, and hence, \cref{eq:shield idle qubits} enforces the shielding of idling qubits---vital to achieving high fidelity.%

Apart from constraints validating the configuration at the beginning of each stage, further constraints are required to relate successive stages.
In an execution stage, only rearranging the qubits is allowed, but no trap transfers are allowed.

\begin{boxedconstraints}[Shuttling in Execution Stage]
\label{con:rydberg transition}
\begin{align}
& \forall_{q,t}\ \ryd{t} \implies \left(\aod{t}{q} = \aod{t + 1}{q}\right) \label{eq:trap invariance}                                                                                                   \\
& \forall_{q,t}\ \ryd{t} \implies \Big(\aod{t}{q} \lor \Big(\Big(\xcoord{t}{q} = \xcoord{t + 1}{q}\Big) \land \Big(\ycoord{t}{q} = \ycoord{t + 1}{q}\Big)\Big)\Big) \label{eq:slm position invariance}  \\
& \forall_{q,t}\ \ryd{t} \implies \Big(\lnot\aod{t}{q} \lor \Big(\Big(\column{t}{q} = \column{t + 1}{q}\Big) \land \Big(\row{t}{q} = \row{t + 1}{q}\Big)\Big)\Big) \label{eq:AOD row/column invariance}
\end{align}
\end{boxedconstraints}

\cref{eq:trap invariance} enforces the invariance of the trap's type.
During shuttling, only qubits in AOD traps can move, cf.~\cref{eq:slm position invariance}, but must stay on their column and row, cf.~\cref{eq:AOD row/column invariance}.

In contrast, transfer stages explicitly allow qubit transfers between trap types. %
First, qubits are stored. %

\begin{boxedconstraints}[Storing in Transfer Stage]
\label{con:transfer slm/store constraints}
\begin{align}
& \forall_{q,t}\ \lnot\ryd{t} \implies \Big(\aod{t+1}{q} \lor \Big(\horizontalOffset{t}{q} = 0 = \verticalOffset{t}{q}\Big)\Big) \label{eq:store in center}                                                         \\
& \forall_{q,t}\ \lnot\ryd{t} \implies \Big(\aod{t+1}{q} \lor \Big(\Big(\xcoord{t}{q} = \xcoord{t+1}{q}\Big) \land \Big(\ycoord{t}{q} = \ycoord{t+1}{q}\Big)\Big)\Big) \label{eq:SLM/store position invariance}     \\
& \forall_{q,t}\ \lnot\ryd{t} \implies \Bigg(\lnot\aod{t}{q} \lor \Bigg(\lnot\aod{t+1}{q} \equiv \storeColumn{t}{\column{t}{q}} \lor \storeRow{t}{\row{t}{q}} \Bigg)\Bigg) \label{eq:at least one store column/row}
\end{align}
\end{boxedconstraints}

The storing can only happen at the location of the SLM trap, \ie, with zero horizontal and vertical offset, cf.~\cref{eq:store in center}.
Next, cf.~\cref{eq:SLM/store position invariance} ensures that stored qubits and those in SLM traps remain at their position.
The transfer of a qubit from an AOD to an SLM trap happens by ramping down the intensity of at least one of the AOD lines crossing the atom.
Consequently, all qubits on this line must be stored, cf.~\cref{eq:at least one store column/row}.

The loading of qubits happens after the storing.
Qubits in AOD traps and those loaded can subsequently shuttle to their next position.
The following \cref{eq:load column topo order invariance} ensures that after the transfer, the relative horizontal order of AOD qubits (loaded ones and those remaining in AOD traps) is maintained.

\begin{boxedconstraints}[Loading and Shuttling in Transfer Stage]
\label{con:transfer load constraints}
\begin{align}
& \begin{aligned}
& \forall_{q,q',t}\: \left(\lnot\ryd{t} \land \aod{t+1}{q} \land \aod{t+1}{q'}\right) \implies \Big(\Big(\xcoord{t}{q} < \xcoord{t}{q'}\Big)\: \lor                                    \\
& \quad \Big(\Big(\xcoord{t}{q} = \xcoord{t}{q'}\Big) \land \Big(\horizontalOffset{t}{q} < \horizontalOffset{t}{q'}\Big)\Big) \equiv \Big(\column{t+1}{q} < \column{t+1}{q'}\Big)\Big)
\end{aligned} \label{eq:load column topo order invariance} %
\end{align}
\end{boxedconstraints}

The vertical order is analogously ensured by a constraint that is omitted for brevity.
Similar to the store operation, \ie,~\cref{eq:at least one store column/row}, the loading operation is constrained similarly, which is also not shown here for brevity.

\subsection{Objective Function}\label{subsec:objective function}
The presented constraints guarantee that satisfying assignments only lead to physically feasible schedules.
However, there is no way to decide which of the possible schedules is the best.
To this end, we introduce an objective function that determines the optimal solution.

As the execution time of a circuit increases, the probability of errors due to decoherence generally rises, resulting in decreased fidelity.
Moreover, each Rydberg beam carries a certain probability of error.
We optimize for both factors by seeking the schedule with the minimum overall number of stages~\(\maxStages\).

\begin{table*}[pt]
    \caption{Layout Comparison}
    \label{tab:results}
    \begin{tabularx}{\linewidth}{l@{\hspace{1ex}}S[table-format=1.0]XS[table-format=3.1]@{\hspace{1ex}}S[table-format=1.0]@{\hspace{1ex}}S[table-format=1.0]@{\hspace{1ex}}S[table-format=1.2]@{\hspace{1ex}}S[table-format=1.2]XS[table-format=3.1]@{\hspace{1ex}}S[table-format=1.0]@{\hspace{1ex}}S[table-format=1.0]@{\hspace{1ex}}S[table-format=1.2]@{\hspace{1ex}}S[table-format=1.2]XS[table-format=3.1]@{\hspace{1ex}}S[table-format=1.0]@{\hspace{1ex}}S[table-format=1.0]@{\hspace{1ex}}S[table-format=1.2]@{\hspace{1ex}}S[table-format=1.2]@{}}
        \toprule
        \textbf{Code}                                  & \textbf{\#CZ} &  & \multicolumn{5}{c}{\textbf{(\ref{layout:no shielding}) No Shielding}} &              & \multicolumn{5}{c}{\textbf{(\ref{layout:bottom storage}) Bottom Storage}} &              & \multicolumn{5}{c}{\textbf{(\ref{layout:double-sided storage}) Double-Sided Storage}}                                                                                                                                                                                                            \\
        &               &  & \faIcon{hourglass}   & \textbf{\#R}                                                          & \textbf{\#T} & \faIcon{clock}                                                            & \textbf{ASP} &                                                                                       & \faIcon{hourglass}  & \textbf{\#R}         & \textbf{\#T}         & \faIcon{clock}            & \textbf{ASP}            &  & \faIcon{hourglass}   & \textbf{\#R}         & \textbf{\#T}         & \faIcon{clock}            & \textbf{ASP}            \\\midrule
        \(\llbracket 7, 1, 3 \rrbracket\) Steane       & 9             &  & <0.1 & 3                                                                     & 0            & 0.35                                                                     & 0.94         &                                                                                       & <0.1 & 3                    & 2                    & 1.56                    & 0.94                    &  & <0.1 & 3                    & 1                    & 1.07                    & 0.94                    \\
        \(\llbracket 9, 1, 3 \rrbracket\) Surface      & 8             &  & <0.1 & 5                                                                     & 0            & 0.52                                                                     & 0.91         &                                                                                       & <0.1 & 3                    & 1                    & 1.15                    & 0.94                    &  & <0.1 & 3                    & 0                    & 0.49                     & 0.95                    \\
        \(\llbracket 9, 1, 3 \rrbracket\) Shor         & 10            &  & <0.1 & 3                                                                     & 0            & 0.66                                                                     & 0.82         &                                                                                       & <0.1 & 5                    & 3                    & 2.33                    & 0.92                    &  & <0.1 & 5                    & 0                    & 0.85                     & 0.94                    \\
        \(\llbracket 15, 7, 3 \rrbracket\) Hamming     & 28            &  & 0.4 & 7                                                                     & 0            & 0.76                                                                     & 0.59         &                                                                                       & 22.5 & 7\textsuperscript{*} & 4\textsuperscript{*} & 2.63\textsuperscript{*} & 0.82\textsuperscript{*} &  & 11.0 & 7\textsuperscript{*} & 2\textsuperscript{*} & 1.94\textsuperscript{*} & 0.83\textsuperscript{*} \\
        \(\llbracket 15, 1, 3 \rrbracket\) Tetrahedral & 28            &  & 0.8 & 7                                                                     & 0            & 0.81                                                                     & 0.59         &                                                                                       & 28.6 & 7\textsuperscript{*} & 4\textsuperscript{*} & 3.01\textsuperscript{*} & 0.82\textsuperscript{*} & & 76.8 & 7\textsuperscript{*} & 2\textsuperscript{*} & 2.18\textsuperscript{*} & 0.83\textsuperscript{*} \\
        \(\llbracket 17, 1, 5 \rrbracket\) Honeycomb   & 32            &  & 0.5 & 7                                                                     & 0            & 0.96                                                                     & 0.57         &                                                                                       & 163.7 & 7\textsuperscript{*} & 5\textsuperscript{*} & 3.00\textsuperscript{*} & 0.80\textsuperscript{*} &  & 313.6 & 7\textsuperscript{*} & 3\textsuperscript{*} & 2.46\textsuperscript{*} & 0.81\textsuperscript{*} \\
        \bottomrule
    \end{tabularx}\vspace{2pt}\\
    \footnotesize
    \faIcon{hourglass}: Comp. time [\SI{}{\hour}] \hfill
    \#CZ: Num. CZ-gates \hfill
    \#R: Num. Rydberg stages \hfill
    \#T: Num. transfer stages \hfill
    \faIcon{clock}: Exec. time [\SI{}{\milli\s}] \hfill
    ASP: Approx. succ. prob.
    \\
    \textsuperscript{*}Results may not be optimal due to solver's timeout after \SI{320}{\hour}.
    \vspace{-8pt}
\end{table*}

\section{Evaluation}\label{sec:evaluation}

The proposed approach is the first that provides a method for optimal state preparation on \emph{zoned} neutral atom architectures.
Experimental evaluations have been conducted to evaluate its applicability.
This section reviews the considered setup and summarizes the obtained results, including a discussion of them.

\subsection{Setting}\label{subsec:setting}
In order to highlight the impact of utilizing the architecture's zones on the overall fidelity%
, we consider three different layouts: %
\begin{enumerate*}[label=(\arabic*), ref=\arabic*]
\item \emph{No Shielding}: one entangling zone without a storage zone, which implies that qubits cannot be shielded and which serves as a baseline\footnote{For this layout, the \cref{eq:shield idle qubits} is unsatisfiable and must be replaced with a constraint ensuring that idling qubits are sufficiently separated, \ie, sit in different interaction sites.};\label[layout]{layout:no shielding}
\item \emph{Bottom Storage}: one storage zone below the entangling zone; and\label[layout]{layout:bottom storage}
\item \emph{Double-Sided Storage}: storage zones on both sides of the entangling zone to see whether the extra storage zone can mitigate the additional shuttling overhead caused by shielding idling qubits\label[layout]{layout:double-sided storage}
\end{enumerate*}.

In all cases, we assume that the entire architecture offers eight columns (\(\maxX\!\!=\!7\)) and seven rows (\(\maxY\!\!=\!6\)), which are split into the zones for the respective layout such that a storage zone always has two rows.
Accordingly, the entangling zone spans all rows in \cref{layout:no shielding} (\(\minEntanglingZone\!\!=\!0,\maxEntanglingZone\!\!=\!6\)), the five top rows in \cref{layout:bottom storage} (\(\minEntanglingZone\!\!=\!2,\maxEntanglingZone\!\!=\!6\)), and the three middle rows in \cref{layout:double-sided storage} (\(\minEntanglingZone\!\!=\!2,\maxEntanglingZone\!\!=\!4\)).
Further, we assume the architecture features six AOD lines in each direction (\(\maxC\!\!=\!\maxR\!\!=\!5\)).
We allow a relative offset in every direction of two (\(\maxH\!\!=\!\!\maxV\!\!=\!2\)).
For qubits to interact, their sites must either be directly or diagonally adjacent (\(\maxInteractionRadius\!=\!2\)).

To meet the objective function, we gradually increment the number of stages \(\maxStages\) until we find a satisfiable instance.
To solve the individual instances we employ the \textsc{Z3}~4.11.2~\cite{demouraZ3EfficientSMT2008} and run it on an Intel\textsuperscript{\textregistered} Xeon\textsuperscript{\textregistered} W-1370P@3.60GHz with 128GB RAM.
The input circuits preparing the \(|0\rangle\)-state of the respective code (similar to circuit in \cref{subfig:motivation circuit}) are generated by taking the stabilizers of popular QEC codes and applying the tool \textsc{Stabgraph}~\cite{amaroScalableCharacterizationLocalizable2020}.

For the assessment of the results, we employ the \emph{Approximated Success Probability}~(ASP)~\cite{schmidComputationalCapabilitiesCompiler2023} as a proxy for the overall fidelity of the circuit.
For a circuit with \(\maxG\) gates, the ASP is defined by %
\begin{align*}
\mathrm{ASP} = \exp\left(-\frac{t_\mathrm{idle}}{T_\mathrm{eff}}\right)\cdot\prod_{i=0}^{\maxG}\mathcal{F}_{g_i}\enspace,
\end{align*}%
where \(t_\mathrm{idle}\) is the accumulated idle time of all qubits, \(T_\mathrm{eff} = \SI{1}{\s}\)~\cite{bluvsteinLogicalQuantumProcessor2023}, and \(\mathcal{F}_{g_i}\) is the fidelity of the respective gate.
The calculation of the ASP is based on the following figures of merit with \(\mathrm{Id_{Ryd}}\) being the faulty identity operation affected by the Rydberg beam:%
\vspace{-3pt}
\par
\begin{center}
\noindent%
\footnotesize%
\newcommand\Shead[1]{\multicolumn{1}{c}{#1}}%
\begin{tabularx}{\linewidth}{XS[table-format=1.5]S[table-format=3.2]S[table-format=1.2]}
\toprule
\textbf{Operation}                                                                                                             & \Shead{\textbf{Fidelity \(\mathcal{F}\)}}          & \Shead{\textbf{Duration \([\SI{}{\micro\s}]\)}} & \Shead{\textbf{Speed \([\SI{}{\micro\s\per\micro\m}]\)}} \\
\midrule
\multicolumn{2}{c}{CZ/\(\mathrm{Id_{Ryd}}\) \hfill 0.995/0.998\cite{everedHighfidelityParallelEntangling2023}\hspace*{-1.8ex}} & 0.27\cite{bluvsteinLogicalQuantumProcessor2023}    & {–}                                                                                                        \\
local RZ                                                                                                                       & 0.99912\cite{bluvsteinLogicalQuantumProcessor2023} & 5\cite{bluvsteinLogicalQuantumProcessor2023}    & {–}                                                      \\
global RY                                                                                                                      & 0.9999\cite{bluvsteinLogicalQuantumProcessor2023}  & 1\cite{bluvsteinLogicalQuantumProcessor2023}    & {–}                                                      \\
Load/Store                                                                                                                     & 0.999\cite{bluvsteinLogicalQuantumProcessor2023}   & 200\cite{bluvsteinLogicalQuantumProcessor2023}  & {–}                                                      \\
Shuttling                                                                                                                      & 1.0\cite{bluvsteinLogicalQuantumProcessor2023}     & {–}                                             & 0.55\cite{bluvsteinQuantumProcessorBased2022}            \\
\bottomrule
\end{tabularx}
\end{center}
With respect to the shuttling distances, we assume that sites within one interaction site are separated by a distance of \(\SI{1}{\micro\m}\), and entire interaction sites by \(\SI{14}{\micro\m}\) to safely avoid long-distance interactions between qubits sitting in neighboring interaction sites~\cite{bluvsteinLogicalQuantumProcessor2023}.
The zones are laid out such that qubits in different zones are at least \(\SI{20}{\micro\m}\) apart from each other~\cite{bluvsteinLogicalQuantumProcessor2023}. %

\subsection{Obtained Results}\label{subsec:results}
\cref{tab:results} summarizes the results.
Here, the first two columns denote the code and the corresponding number of \mbox{CZ-gates} needed to prepare the state.
Then, the following columns provide the obtained results: \faIcon{hourglass} denotes the time taken by Z3 to solve the particular instance, \#R and \#T denote the number of Rydberg and transfer stages, respectively, \faIcon{clock} denotes the overall execution time it takes to perform the schedule on the neutral atom architecture, and ASP denotes the approximated success probability.
The latter is the primary measure to compare different schedules.
For \cref{layout:bottom storage,layout:double-sided storage}, the difference in the ASP compared to \cref{layout:no shielding} is highlighted in \cref{fig:plot_est}.
The higher the bar is, the more significant the respective improvement.
Note that---despite the complexity---for the smaller codes, \ie, the first three, the solver terminates within hundreds of milliseconds; for the larger codes, however, it takes up to several days for individual SMT instances and in few cases we obtain results not guaranteed to be optimal due to timeout.

\subsection{Discussion}\label{ref:discussion}

The results confirm that the proposed approach---for the first time---can generate minimal schedules for state preparation on the zoned neutral atom architecture.
While the approach has scalability limitations, as expected due to the problem's inherent complexity, this is not a significant concern: the schedules only need to be generated once and can serve as optimal building blocks that are instantly usable during compilation.

The proposed approach also allows to evaluate the benefits of the \emph{zoned} neutral atom architecture for the first time.
In fact, \cref{tab:results} reveals consistently higher fidelities, reflected by the ASP, for \cref{layout:bottom storage,layout:double-sided storage} compared to \cref{layout:no shielding}, which clearly shows the advantage of shielding idling qubits in the storage zone.
The proposed approach achieves this advantage even in the cases where the results for \cref{layout:bottom storage,layout:double-sided storage} are not guaranteed to be optimal.
Existing solutions do not account for zones, with all operations occurring implicitly in the entangling zone.
As a result, their performance can only match, but never exceed, the figures presented for \cref{layout:no shielding}.

\begin{figure}[hp]
\vspace{-6pt}
\centering
\includegraphics[width=\linewidth]{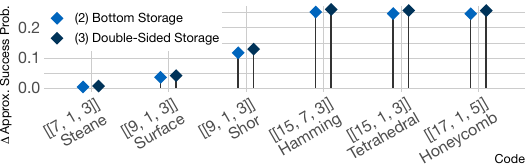}\\
\vspace{-8pt}
\caption{Differences in the ASP for different codes and layouts.}
\label{fig:plot_est}
\vspace{-8pt}
\end{figure}

Moreover, \cref{fig:plot_est} shows that the ASP improves slightly further by having storage zones on both sides of the entangling zone since shuttling distances are shorter and fewer transfer stages are required.
These results provide valuable insights for the design of future quantum devices.

\section{Conclusions}\label{sec:conclusions}

Zoned neutral atom architectures offer an excellent opportunity to shield idle qubits from interfering influences.
However, so far, those architectures lack appropriate software support.
In this work, we have proposed an SMT-based approach to generate minimal schedules for state preparation circuits tailored to this architecture.
The evaluation demonstrates the benefits of the proposed technique for shielding idling qubits and for exploring different hardware design choices.

\subsection*{Acknowledgements}\label{sec:ack}
\footnotesize
This work received funding from the European Research Council (ERC) under the European Union’s Horizon 2020 research and innovation program (grant agreement No. 101001318), was part of the Munich Quantum Valley, which the Bavarian state government supports with funds from the Hightech Agenda Bayern Plus, and has been supported by the BMWK based on a decision by the German Bundestag through project QuaST, as well as by the BMK, BMDW, and the State of Upper Austria in the frame of the COMET program (managed by the FFG).
\clearpage
\enlargethispage{-27ex}
\printbibliography

\end{document}

\typeout{get arXiv to do 4 passes: Label(s) may have changed. Rerun}